\documentclass[fleqn,10pt]{wlscirep}
\usepackage[utf8]{inputenc}
\usepackage[T1]{fontenc}
\usepackage{graphicx}
\usepackage{epstopdf}
\usepackage{color}
\title{Electrostatic wave interaction via asymmetric vector solitons
as precursor to rogue wave formation in non-Maxwellian plasmas}
\author[1,*]{N. Lazarides}
\author[2]{Giorgos P. Veldes}
\author[3]{D. J. Frantzeskakis}
\author[1,4,3,5]{Ioannis Kourakis}
\affil[1]{
Khalifa University of Science and Technology, Department of Mathematics, 
Abu Dhabi, P.O. Box 127788, United Arab Emirates}
\affil[2]{
University of Thessaly, Department of Physics, Lamia 35100, Greece}
\affil[3]{Department of Physics, National and Kapodistrian University of Athens,
GR-15784 Zografou, Athens, Greece}
\affil[4]{Space $\&$ Planetary Science Center, Khalifa University of Science and Technology, 
Abu Dhabi, P. O. Box 127788,  United Arab Emirates}
\affil[5]{Hellenic Space Center, Leoforos Kifissias 178, Chalandri, Athens, GR-15231, Greece}
\affil[*]{nikolaos.lazarides.1966@gmail.com}

\begin{abstract}
An asymmetric pair of coupled nonlinear Schr{\"o}dinger (CNLS) equations has been 
derived through a multiscale perturbation method applied to a plasma fluid 
model, in which two wavepackets of distinct (carrier) wavenumbers ($k_1$ and $k_2$) 
and amplitudes ($\Psi_1$ and $\Psi_2$) are allowed to co-propagate and interact. 
The original fluid model was set up for a non-magnetized plasma consisting of 
cold inertial ions evolving against a $\kappa-$distributed electron 
background in one dimension. The reduction procedure resulting in the 
CNLS equations has provided analytical expressions for the dispersion, 
self-modulation and cross-coupling coefficients in terms  of
the two carrier wavenumbers. These coefficients present no symmetry whatsoever, 
in the general case (of different wavenumbers).

The possibility for coupled envelope (vector soliton) solutions to occur has been 
investigated. Although the CNLS equations are asymmetric and non-integrable, 
in principle, the system admits various types of vector soliton solutions,  
physically representing nonlinear, localized electrostatic plasma modes, whose 
areas of existence is calculated on the wavenumbers' parameter plane. 
The possibility for either bright (B) or dark (D) type excitations for either of 
the (2) waves provides four (4) combinations for the envelope pair (BB, BD, DB, DD), 
if a set of explicit criteria is satisfied. Moreover, the soliton parameters 
(maximum  amplitude, width) are also calculated for each type of vector soliton 
solution, in its respective area of existence. \color{black}
The dependence of the vector soliton characteristics on the (two) carrier wavenumbers 
and on the spectral index $\kappa$ 
characterizing the electron distribution has been explored.   %
In certain cases, the (envelope) amplitude of one component may exceed its 
counterpart (second amplitude) by a factor 2.5 or higher, indicating that
extremely asymmetric waves may be formed due to modulational interactions 
among copropagating wavepackets.

As $\kappa$ decreases from large values, modulational instability occurs in 
larger areas of the parameter plane(s) and with higher growth rates. 
The distribution of different types of vector solitons on the parameter plane(s) 
also varies significantly with decreasing 
$\kappa$, and in fact dramatically for $\kappa$ between $3$ 
and $2$. Deviation from the Maxwell-Boltzmann picture therefore seems to favor 
modulational instability as a precursor to the formation of bright 
(predominantly) type envelope excitations and freak waves. 
\end{abstract}

\begin{document}
\flushbottom
\maketitle
\thispagestyle{empty}

\section{Introduction}

A pair of nonlinearly coupled nonlinear Schr{\"o}dinger equations (hereafter 
referred to as the CNLS system of equations) arises as a prototype model of 
mathematical physics, which occurs in various physical contexts 
\cite{Leble2009,Charalampidis2015,Kevrekidis2016,Stalin2021}, including 
water waves \cite{He2022,Ablowitz2015}, 
left-handed (negative refraction index) transmission lines \cite{Veldes2013}, 
optical pulse propagation in birefringence fibers
\cite{Menyuk1987,Frisquet2016,Huang2022},
and in optical nonlinear media \cite{Kivshar1993}, 
vector solitons in left-handed metamaterials \cite{Lazarides2005},
polarized pulse pair propagation in anisotropic dispersive 
media \cite{Tyutin2022},
in electrically driven graphene multilayer mediums \cite{Shaukat2022},
pulse propagation in isotropic Kerr media with chromatic dispersion 
\cite{Haelterman1994},
and even breathers and rogue waves in optical fibers \cite{Wu2023}. 
Formally similar systems of equations have been used to model light (beam) 
propagation 
\cite{McKinstrie1989,McKinstrie1990,Luther1990,Luther1992} and electrostatic 
or electromagnetic wave propagation in plasmas  
\cite{Spatschek1978,Som1979,Kourakis2005b,Singh2013,Borhanian2017,Kofane2020,Lazarides2023a,Lazarides2023b}.
Independently from a physical context, various studies of vector solitons and 
rogue waves have been carried out, based on general CNLS models 
\cite{Buryak1996,Guo2011,Baronio2012,He2014,Ablowitz2015,Li2015,Nath2017}, and 
variants of CNLS equations such as coupled derivative 
nonlinear Schr{\"o}dinger equations \cite{Xiang2022,Jin2023}, vector 
($N-$component) CNLS \cite{Jiang2012,Yang2022,Liu2022}, 
nonlocal CNLS \cite{Ren2022}, 
CNLS equations with variable coefficients \cite{Yu2014}, 
coherently coupled CNLS equations \cite{Zhang2018}, and 
coupled high-order nonlinear Schr{\"o}dinger equations \cite{Zhou2022}, among 
others, have been investigated with respect to vector solitons and rogue waves.

A pair of nonlinearly coupled CNLS equations was recently 
\cite{Lazarides2023a,Lazarides2023b} derived from a plasma model consisting of 
a cold inertial ion fluid evolving against an electron background. Generalizing 
an earlier study involving a thermal (Maxwellian) electron distribution 
\cite{Lazarides2023a}, the formalism has been recently \cite{Lazarides2023b} 
adopted to electron population(s) that follows a kappa distribution 
\cite{Livadiotis2013,Livadiotis2015b,Livadiotis2019,Nicolaou-Livadiotis2020,
Elkamash2021,Saberian2022,Mukherjee2022,Hatami2022}. 
The kappa (family of) distribution functions (DF) is characterized by a spectral 
index $\kappa$, and exhibits a high-energy tail in the large (suprathermal) 
range of electron velocities. The analytical expression of the kappa DF 
converges to the Maxwell-Boltzmann distribution for infinite $\kappa$. 
Such distributions are a common occurrence in Space plasma observations,  
e.g. in the solar wind \cite{Livadiotis2013,Livadiotis2015b,Livadiotis2017} 
and in planetary magnetospheres \cite{SSVIKSciRep}. 
The original plasma fluid  model and its lengthy algebraic reduction to a pair 
of CNLS equations for the envelopes of two modulated electrostatic wavepackets, 
by  using a Newell type multi-scale perturbation technique,  
is described in great detail in Refs. \cite{Lazarides2023a,Lazarides2023b}, so 
the details will be omitted in the following. The main outcome of that study, 
in the form of the (six) coefficients involved in the resulting CNLS equations, 
will be presented here, for completeness, in terms of the wavenumbers of the 
two interacting waves and the spectral index $\kappa$ which characterizes the 
electron distribution.  
Based on the system of CNLS equations obtained in those earlier studies, our 
ambition in the paper at hand is to investigate the existence of coupled 
localized envelope modes (vector solitons), from first principles, and to 
explore their dependence on the intrinsic plasma parameters, namely the 
two wavenumbers ($k_1$, $k_2$) and $\kappa$. 

It must be pointed out that the CNLS system of equations that forms the basis 
of our study is \emph{not} amenable to the widely studied (integrable) 
Manakov system \cite{Manakov1974}, 
unless identical carrier waves ($k_1=k_2$) are considered. Indeed, for arbitrary 
wavenumbers of the two co-propagating waves, the coefficients of the CNLS 
equations do not exhibit any known symmetry,  hence the system is rendered 
non-symmetric. For the same reason, any attempt for reduction of the number of 
the coefficients of these CNLS equations cannot give less than four coefficients 
\cite{Tan-Boyd2000,Tan-Boyd2001}.

The CNLS system of equations in its general form admits vector soliton solutions, 
rogue waves, and breathers, which can be obtained either analytically or 
numerically. In this paper, within the context of the plasma fluid model 
considered for electrostatic waves, we have obtained four different types of 
vector solitons, whose components are actually combinations of bright and dark 
type envelope solitons, i.e. reminiscent of solutions of the single nonlinear 
Schr{\"o}dinger equation. We have derived a set of analytical conditions for 
the existence of such vector solitons, in terms of the various coefficients 
(assumed to take arbitrary values), and we have subsequently explored their 
parametric dependence on the relevant plasma parameters: the two wavevectors and 
the spectral index $\kappa$. In each of these existence regions, on the 
($k_1,k_2$) plane, we have also calculated the vector soliton parameters, i.e., 
the envelope amplitudes and their (common) width. These  (amplitude and width)  
obviously vary, upon a variation of either of the wavenumbers or the spectral 
index $\kappa$. Several illustrative examples are shown in the following,  
in which a structural transition between a particular type of a vector soliton 
(and its parameters) and another can be observed. These transition may either be 
smooth, or take place through a divergence of the width and the amplitudes at a 
transition point (boundary between different regions). In certain cases, one of 
the components of a vector soliton may acquire a very high amplitude with respect 
to that of its sister component. We may characterize these solutions as extremely
asymmetric vector solitons, a configuration which has not been discussed before.


\section{Asymmetric Nonlinear Schr{\"o}dinger Equations and Coefficients}

A plasma fluid model was considered in earlier work \cite{Lazarides2023b}, as a 
basis to describe electrostatic (ion-acoustic) waves. A non-magnetized plasma 
was considered, consisting of a cold inertial ion fluid evolving against a 
``thermalized" (highly energetic) electron background, in a one-dimensional 
geometry. Given the large mass disparity between the electrons and the massive 
ions, the former were assumed to be inertia-less, thus characterized by  an 
equilibrium configuration, modeled as a $\kappa-$type velocity distribution. 
Two co-propagating wavepackets were considered, with wavevectors $k_1$ and $k_2$, 
respectively. A Newell type multiple scale perturbation technique led, after a 
tedious calculation \cite{Lazarides2023b}, to the following 
pair of CNLS equations:  
\begin{eqnarray}
\label{eq10}
   i \left( \frac{\partial \Psi_1}{\partial t_2} 
    +v_{g,1} \frac{\partial \Psi_1}{\partial x_2} \right)
   +P_1 \frac{\partial^2 \Psi_1}{\partial x_1^2}
   +\left(Q_{11} |\Psi_1|^2 +Q_{12} |\Psi_2|^2 \right) \Psi_1 =0, 
\\
\label{eq11}
   i \left( \frac{\partial \Psi_2}{\partial t_2} 
    +v_{g,2} \frac{\partial \Psi_2}{\partial x_2} \right)
    +P_2 \frac{\partial^2 \Psi_2}{\partial x_1^2}
    +\left(Q_{21} |\Psi_1|^2 +Q_{22} |\Psi_2|^2 \right) \Psi_2 =0 \, , 
\end{eqnarray}
where
\begin{equation}
\label{eq12}
   P_j =-\frac{3}{2} \frac{c_1 k_j}{(k_j^2 +c_1)^{5/2}} = 
          \frac{1}{2} \frac{\partial^2 \omega_j}{\partial k_j^2}, 
          \qquad
  v_{g,j} =\frac{c_1}{(k_j^2 +c_1)^{3/2}} =\frac{\partial \omega_j}{\partial k_j}, 
                  \qquad
   \omega_j =\frac{k_j}{\sqrt{k_j^2 +c_1}}
\end{equation} 
are the (linear) dispersion coefficients, the group velocities, and the frequency 
dispersion relations, respectively.  Note that $P_j = \frac{1}{2}\frac{\partial^2 \omega_j}{\partial k_j^2}$, 
in a way formally analogous to the group-velocity-dispersion (GVD) terms known in nonlinear optics. 
$Q_{11}$ and $Q_{22}$ are self-modulation coefficients and $Q_{12}$ and $Q_{21}$ 
are cross-coupling coefficients. 
For the exact expressions providing the (four) coefficients $Q_{ij}$ 
(for $i, j = $ 1 or 2) as functions of $k_1$, $k_2$ and $\kappa$ 
(via the constants $c_1$, $c_2$, and $c_3$), see the Supplementary Information part accompanying this article.
We emphasize that, unless $k_1 = k_2$ (a stringent condition that won't be satisfied, 
in general), the above pairs of coefficients take different values, viz. $P_1 \ne P_2$, 
$Q_{11} \ne Q_{22}$ and $Q_{12} \ne Q_{21}$.

The constants $c_1$, $c_2$, and $c_3$, incorporating the effect of $\kappa$, are given as  by
\cite{Baluku2010}
\begin{eqnarray}
\label{eq15}
   c_1 =\frac{\kappa -\frac{1}{2}}{\kappa -\frac{3}{2}},  
 \qquad
   c_2 =\frac{\left(\kappa -\frac{1}{2}\right) \left(\kappa +\frac{1}{2}\right)}
             {2! \left(\kappa -\frac{3}{2} \right)^2}, 
\qquad
   c_3 =\frac{\left(\kappa -\frac{1}{2}\right) \left(\kappa 
       +\frac{1}{2}\right) \left(\kappa +\frac{3}{2}\right)}
             {3! \left(\kappa -\frac{3}{2}\right)^3}. \label{cjdef}
\end{eqnarray}
The above expressions are the first three coefficients in the Mc Laurin expansion 
$n_e \simeq 1 +c_1 \phi +c_2 \phi^2 +c_3 \phi^3 + \cdots$, of the electron number density 
\begin{equation}
\label{eq18}
   n_e =\left( 1 -\frac{\phi}{\kappa -\frac{3}{2}} \right)^{-\left(\kappa -\frac{1}{2} 
         \right)},
\end{equation}
that is obtained upon integrating the kappa distribution 
\cite{Livadiotis2013,Livadiotis-McComas2009,Livadiotis2017}, actually a 
straightforward replacement for the Maxwell-Boltzmann distribution when dealing 
space and astrophysical plasmas \cite{Nicolaou-Livadiotis2020,Nicolaou-Livadiotis2018}. 
The nature and composition of the plasmas allows for extracting the spectral 
index $\kappa$ in each particular case. Indeed, kappa distributions with 
$2 < \kappa < 6$ have been found to fit the observations and satellite data in 
the solar wind \cite{Pierrard-Lazar2010}, among other successful examples of 
data fitting in space plasmas, making the kappa distribution an ubiquitous 
paradigm in Space science \cite{Livadiotis2017}.

The CNLS equations (\ref{eq10}) and (\ref{eq11}) can be transformed to 
\begin{eqnarray}
\label{eq19}
   i \left( \frac{\partial \Psi_1}{\partial \tau} 
   +\delta \frac{\partial \Psi_1}{\partial \xi} \right)
   +P_1 \frac{\partial^2 \Psi_1}{\partial \xi^2}
   +\left(Q_{11} |\Psi_1|^2 +Q_{12} |\Psi_2|^2 \right) \Psi_1 =0, 
\\
\label{eq20}
   i \left( \frac{\partial \Psi_2}{\partial \tau} 
   -\delta \frac{\partial \Psi_2}{\partial \xi} \right)
   +P_2 \frac{\partial^2 \Psi_2}{\partial \xi^2}
   +\left(Q_{21} |\Psi_1|^2 +Q_{22} |\Psi_2|^2 \right) \Psi_2 =0,
\end{eqnarray}
after a change of the independent variables $x$ and $t$ (in which the 
subscripts have been dropped) through $\xi = x -v t$ and $\tau =t$, with
$v =( v_{g,1} +v_{g,2} )/2$ and $\delta =( v_{g,1} -v_{g,2} )/2$ being
is the half-sum and the half-difference of the group velocities, 
respectively.

Note that the ``walk-off'' parameter $\delta$ is, by its definition, clearly a 
function of both $k_1$ and $k_2$ (in addition to $\kappa$), through the group velocities $v_{g,j}$
($j=1,2$) of the two interacting wavepackets in the plasma. 
Generally speaking, a large $\delta$ would be able to cause dynamic 
instabilities in the system and eventually prevent the formation of 
various types of vector solitons (to be discussed later in this article). However, from the expression of
the group velocities $v_{g,j}$ in Eq. (\ref{eq12}), we may observe that 
their values are limited in the interval $[- 1/\sqrt{c_1}, + 1/\sqrt{c_1}]$, where $c_1 =c_1(\kappa)$ -- 
defined in (\ref{cjdef}) above -- exceeds unity. 
It follow that, in the Maxwellian case ($\kappa =100$), $c_1 =1$ and the extremal 
values of $\delta$ are $\pm 0.5$ (assuming co-propagating wavepackets), while for lower values of $\kappa$, 
the corresponding extremal values of $\delta$ are (in absolute value) even smaller, 
all the way down to zero (attained for $\kappa = 3/2$).  
Therefore, the walk-off parameter acquires small values,  and is not expected to prevent soliton
formation.

Then, by applying the transformation
$\Psi_1 = \bar{\Psi}_1\, \exp\left[ i \left( \frac{\delta^2}{4 P_1} \tau -\frac{\delta}{2 P_1} \xi \right) \right]$
and 
$\Psi_2 = \bar{\Psi}_2\, \exp\left[ i \left( \frac{\delta^2}{4 P_2} \tau +\frac{\delta}{2 P_2} \xi \right) \right]$
to Eqs. (\ref{eq19}) and (\ref{eq20}) we obtain the more familiar form
\begin{eqnarray}
\label{eq22}
   i \frac{\partial \bar{\Psi}_1}{\partial \tau} 
   +P_1 \frac{\partial^2 \bar{\Psi}_1}{\partial \xi^2}
   +\left(Q_{11} |\bar{\Psi}_1|^2 +Q_{12} |\bar{\Psi}_2|^2 \right) \bar{\Psi}_1 =0, 
\\
\label{eq23}
   i \frac{\partial \bar{\Psi}_2}{\partial \tau} 
   +P_2 \frac{\partial^2 \bar{\Psi}_2}{\partial \xi^2}
   +\left(Q_{21} |\bar{\Psi}_1|^2 +Q_{22} |\bar{\Psi}_2|^2 \right) \bar{\Psi}_2 =0,
\end{eqnarray}
where $\bar{\Psi}_j$ are complex functions of the new variables $\xi$ and $\tau$.

\section{Modulational Instability: Compatibility Condition and Growth Rate}

Modulational instability (MI) analysis for two co-propagating plane-wave 
solutions of the CNLS equations (\ref{eq19}) and (\ref{eq20}) can be performed 
using the procedure described e.g., in Refs. 
\cite{Kourakis2006}. The plane waves 
\begin{equation}
\label{mi01}
   \Psi_j =\Psi_{j,0} \, e^{i \tilde{\omega}_j \tau},
\end{equation}
where $\Psi_{j,0}$ ($j=1,2$) is a constant real amplitude and $\tilde{\omega}_j$ 
is an internal frequency, are solutions of the CNLS equations for 
$\tilde{\omega}_1 =Q_{11} \Psi_{1,0}^2 +Q_{12} \Psi_{2,0}^2$ and
$\tilde{\omega}_2 =Q_{21} \Psi_{1,0}^2 +Q_{22} \Psi_{2,0}^2$,
and constitute nonlinear modes that may become modulationally unstable in the 
presence of a small amplitude perturbation of wavenumber $K$ and perturbation 
frequency $\Omega$. By following the standard approach, we find that the above 
nonlinear modes are unstable whenever the {\em compatibility condition} 
\begin{equation}
\label{mi10}
   \left[ (\Omega -\delta K)^2 -\Omega_1^2 \right] 
   \left[ (\Omega +\delta K)^2 -\Omega_2^2 \right] =\Omega_c^4,
\end{equation}
where $\Omega_j^2 =P_j K^2 \left( P_j K^2 -2 Q_{jj} \Psi_{j,0}^2 \right)$ 
and $\Omega_c^4 =4 P_1 P_2 Q_{12} Q_{21} \Psi_{1,0}^2 \Psi_{2,0}^2 K^4$,
has at least one pair of complex conjugate roots ($j=1,2$). Then, the (positive) 
imaginary part of these roots provides the {\em growth rate} $\Gamma$ of the 
modulationally unstable modes. When Eq. (\ref{mi10}) has two pairs of complex 
conjugate roots, then the largest of their imaginary parts provide the growth 
rate of the modulationally unstable modes, i.e.,
\begin{equation}
\label{mi13}
   \Gamma ={\rm max}\{{\rm Im}(\Omega_r)\},       
\end{equation}
where $\Omega_r$ denotes one of the four roots of the compatibility condition
above.

The growth rate $\Gamma$ is calculated numerically for the CNLS equations by 
finding the roots of the fourth degree (in $\Omega$) polynomial resulting from 
the compatibility condition, Eq. (\ref{mi10}). Then $\Gamma$ is mapped on the
$k_1 - k_2$ parameter plane for several values of the perturbation wavenumber 
$K$, and two values of the spectral index $\kappa$, i.e., for $\kappa =2$ and 
$\kappa =3$. The amplitudes of the nonlinear wave modes in these calculations
are $\Psi_{1,0} =\Psi_{2,0} =0.1$.

\begin{figure}[htp]
    \centering
    \includegraphics[width=16cm]{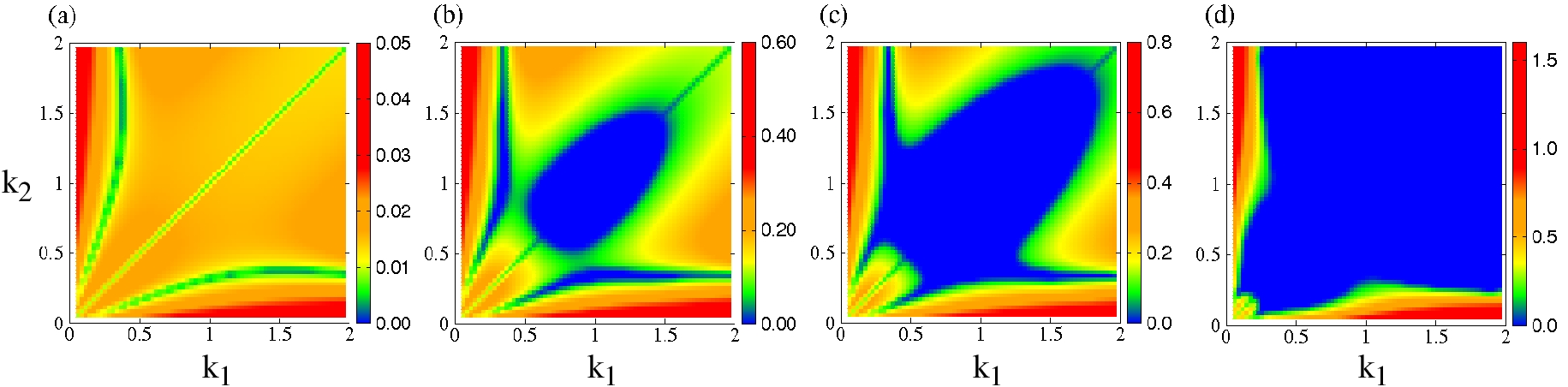}
    \caption{
    Maps of the growth rate $\Gamma$ on the $k_1 - k_2$ plane for $\kappa =2$,
    $\Psi_{1,0} =\Psi_{2,0} =0.1$, and perturbation wavenumber (a) $K =0.1$;
    (b) $K =1.3$; (c) $K =1.6$; (d) $K =3.4$. The values of $K$ are chosen so 
    that they illustrate the variability of the growth rate patterns in the 
    best possible way. 
    }
    \label{fig1}
\end{figure}
\begin{figure}[htp]
    \centering
    \includegraphics[width=16cm]{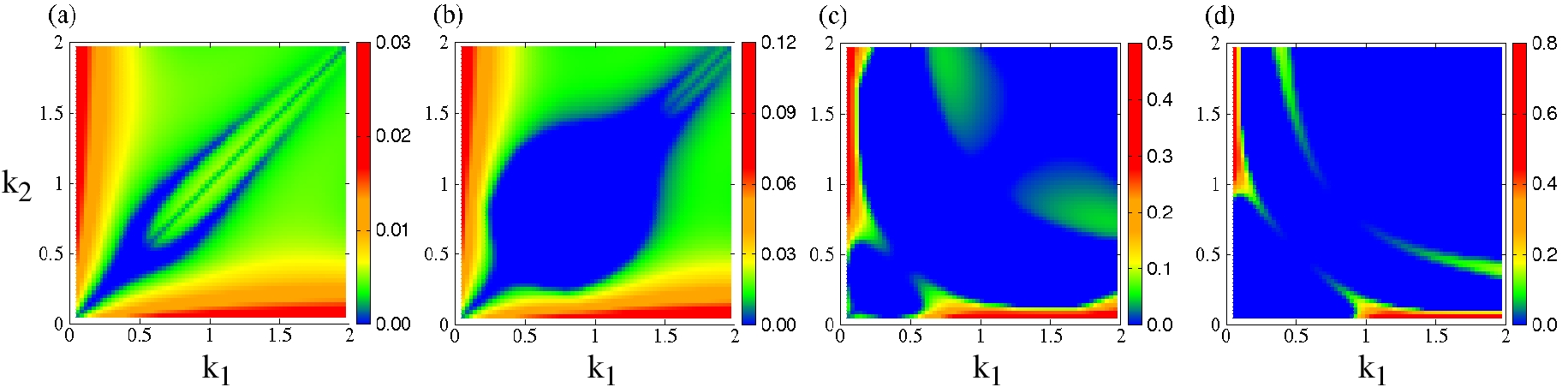}
    \caption{
    Maps of the growth rate $\Gamma$ on the $k_1 - k_2$ plane for $\kappa =3$,
    $\Psi_{1,0} =\Psi_{2,0} =0.1$, and perturbation wavenumber (a) $K =0.1$;
    (b) $K =0.4$; (c) $K =1.9$; (d) $K =3.4$. 
    The values of $K$ are chosen so that they illustrate the variability of the 
    growth rate patterns in the best possible way. 
    }
    \label{fig2}
\end{figure}

In Figs. \ref{fig1} and \ref{fig2}, maps of the growth rate $\Gamma$ are shown 
on the $k_1 - k_2$ parameter plane of the wavenumbers of the two co-propagating
(carrier) waves for spectral index $\kappa =2$ and $\kappa =3$, respectively. 
These values of $\kappa$ were selected to be into the range of 
physically acceptable values for space plasmas that have been observed to be
in the interval from $\kappa =2$ to $\kappa =6$. Moreover, they are more or 
less symmetrically arranged around the value $\kappa \simeq 2.5$. 
In Fig. \ref{fig1}, in particular, for $\kappa =2$, the system is modulationally 
unstable for low values of the perturbation wavenumber $K$, for all the parameter 
plane shown (Fig. \ref{fig1}(a)). At around $K \approx 1.2$, the first 
modulationally stable island becomes visible, which acquires substantial area 
roughly in the middle of the $k_1 - k_2$ parameter plane for $K=1.3$ 
(blue color, Fig. \ref{fig1}(c)). Note that there are also two narrow 
modulationally stable (blue) areas develop around the curves where $Q_{12}$ and 
$Q_{21}$ are zero. For $K$ larger than $1.3$, the modulationally stable areas 
grow and merge together into a single large one, which continues 
to grow with increasing $K$ (Figs. \ref{fig1}(c)-(d)). For $K$ larger than $3.4$,
almost all of the $k_1 - k_2$ parameter plane shown in Fig. \ref{fig1} becomes
modulationally stable. Note the strong modulational instability in 
Fig. \ref{fig1}(d) that is limited at low $k_1$ and low $k_2$, where the nonlinear 
coupling coefficients $Q_{12}$ and $Q_{21}$ acquire very large values.

In Fig. \ref{fig2}, for $\kappa =3$, the system possess a modulationally stable 
area even for low values of $K$, as e.g., can be observed in Fig. \ref{fig2}(a)
(blue color), which is close and around the diagonal $k_1 =k_2$. Already at 
$K =0.4$, the modulationally unstable area has grown significantly 
(Fig. \ref{fig2}(b)), while it continues to grow with increasing value of $K$
(Figs. \ref{fig2}(c)-(d)). For $K$ larger than $3.4$ almost all of the parameter 
plane shown is modulationally stable. Again, the strongly unstable areas are 
limited in the low $k_1$ and low $k_2$ areas of the plane.   

\section{Vector Solitons: Existence and their Parameters}

The CNLS equations Eqs. (\ref{eq22}) and (\ref{eq23}) admit several types of 
vector soliton solutions that are combinations of bright and dark soliton 
solutions of the single NLS equation. Indeed, as we shall show below, four types 
of vector solitons, i.e., bright-bright (BB), bright-dark (BD), dark-bright (DB), 
ans dark-dark (DD), may exist in the CNLS system Eqs. (\ref{eq22}) and 
(\ref{eq23}). Their parameters, i.e., their amplitudes $A_1$ and $A_2$, their 
(common) width $b$, and their internal frequencies $\nu_1$ and $\nu_2$ are related 
through simple mathematical expressions to the coefficients $P_j$ and $Q_{ij}$ of 
the CNLS equations (\ref{eq22}) and (\ref{eq23}). Each component of these vector 
solitons represent modulated electrostatic wavepackets which are moving in the 
plasma  and interact strongly and
nonlinearly with strengths $Q_{12}$ and $Q_{21}$.

The four types of vector solitons that may exist in the system of CNLS equations
are given below along with their parameters, i.e., their amplitudes $A_1$ and $A_2$
and their width $b$, as a function of the CNLS coefficients $P_j$ and $Q_{ij}$
($j=1,2$) which are in turn functions of the wavenumbers $k_1$ and $k_2$. Note 
that in the vector solitons expressions below one of the parameters may take 
arbitrary values. Here, we choose the amplitude $A_1$ as the free parameter and 
fix it to $A_1 =0.1$ in what follows.
\\

\noindent {\em Case I: Bright-bright (BB) vector solitons.} \\
We seek for bright-bright (BB) vector solitons in the form  
\begin{eqnarray}
\label{sl01}
    \bar{\Psi}_1 =A_1\, {\rm sech}\left( b \xi \right) e^{-i \nu_1 \tau}, 
\qquad
    \bar{\Psi}_2 =A_2\, {\rm sech}\left( b \xi \right) e^{-i \nu_2 \tau}. 
\end{eqnarray}
By substitution of Eqs. (\ref{sl01}) into Eqs. (\ref{eq22}) and (\ref{eq23}) we obtain
\begin{equation}
\label{sl03}
    \left( {A_2}/{A_1} \right)^2 =-\alpha, \qquad \left( {b}/{A_1} \right)^2 =-\beta,
\end{equation}
and $\nu_1 =-b^2 P_1$, $\nu_2=-b^2 P_2$, where 
\begin{eqnarray}
\label{sl05}
    \alpha =\frac{Q_{21} P_1 -Q_{11} P_2}{P_1 Q_{22} -P_2 Q_{12}},
\qquad
    \beta =\frac{1}{2} \frac{Q_{11} Q_{22} -Q_{21} Q_{12}}{P_2 Q_{12} -P_1 Q_{22}}.
\end{eqnarray}
\\

\noindent {\em Case II: Bright-dark (BD) vector solitons.} \\
We seek for bright-dark (BB) vector solitons in the form  
\begin{eqnarray}
\label{sl07}
    \bar{\Psi}_1 =A_1\, {\rm sech}\left( b \xi \right) e^{-i \nu_1 \tau}, 
\qquad
    \bar{\Psi}_2 =A_2\, \tanh\left( b \xi \right) e^{-i \nu_2 \tau}. 
\end{eqnarray}
By substitution of Eqs. (\ref{sl07}) into Eqs. (\ref{eq22}) and (\ref{eq23}) 
we obtain
\begin{equation}
\label{sl09}
     \left( {A_2}/{A_1} \right)^2 =+\alpha, \qquad \left( {b}/{A_1} \right)^2 =-\beta, 
\end{equation}
and $\nu_1 =-b^2 P_1 -Q_{12} A_2^2$, $\nu_2=-Q_{22} A_2^2$, where $\alpha$ 
and $\beta$ are given by Eqs. (\ref{sl05}).
\\

\noindent {\em Case III: Dark-bright (DB) vector solitons.} \\
We seek for dark-bright (DB) vector solitons in the form  
\begin{eqnarray}
\label{sl11}
    \bar{\Psi}_1 =A_1\, \tanh\left( b \xi \right) e^{-i \nu_1 \tau}, 
\qquad
    \bar{\Psi}_2 =A_2\, {\rm sech}\left( b \xi \right) e^{-i \nu_2 \tau}. 
\end{eqnarray}
By substitution of Eqs. (\ref{sl11}) into Eqs. (\ref{eq22}) 
and (\ref{eq23}) we obtain
\begin{equation}
\label{sl13}
    \left( {A_2}/{A_1} \right)^2 =+\alpha, \qquad \left( {b}/{A_1} \right)^2 =+\beta,
\end{equation}
and $\nu_1 = - \left( Q_{11} A_1^2 +Q_{12} A_2^2\right)$,
$\nu_2= - \left(b^2 P_2 +Q_{21} A_1^2\right)$,
where $\alpha$ and $\beta$ are given by Eqs. (\ref{sl05}).
\\

\noindent {\em Case IV: Dark-dark (DD) vector solitons.} 

We seek for dark-bright (DB) vector solitons in the form  
\begin{eqnarray}
\label{sl15}
    \bar{\Psi}_1 =A_1\, \tanh\left( b \xi \right) e^{-i \nu_1 \tau}, 
 \qquad
    \bar{\Psi}_2 =A_2\, \tanh\left( b \xi \right) e^{-i \nu_2 \tau}. 
\end{eqnarray}
By substitution of Eqs. (\ref{sl15}) into Eqs. (\ref{eq22}) and (\ref{eq23}) we obtain
\begin{equation}
\label{sl17}
    \left( {A_2}/{A_1} \right)^2 =-\alpha, \qquad \left( {b}/{A_1} \right)^2 =+\beta, 
\end{equation}
and $\nu_1 =- \left(Q_{11} A_1^2 +Q_{12} A_2^2 \right)$,
$\nu_2 =-\left( Q_{21} A_1^2 +Q_{22} A_2^2 \right)$,
where $\alpha$ and $\beta$ are given by Eqs. (\ref{sl05}).
\\

Thus, the vector soliton parameters $A_j$, $b$, and $\nu_j$ are actually 
themselves functions of the wavenumbers $k_1$ and $k_2$ of the two 
co-propagating wavepackets through the CNLS coefficients $P_j$ and $Q_{ij}$ 
($j=1,2$). The simplest way to calculate them is to first fix one of them 
(one of the amplitudes selected as reference, say $A_1$), and then calculate $A_2$ and $b$ 
(e.g., from Eqs. (\ref{sl17}) in Case IV), and last the frequencies $\nu_j$. 

For BB, BD, DB and DD vector solitons to exist in the CNLS system that we 
have obtain from the particular plasma fluid model considered here, the 
right-hand-sides of Eqs. (\ref{sl03}), (\ref{sl09}), (\ref{sl13}), and 
(\ref{sl17}), respectively, must be greater than zero (so that $A_2$ and $b$ 
are real; $A_1$ is of course fixed to a real number). This condition for the 
existence of the four types of vector solitons can be expressed simply as
\begin{eqnarray}
\label{sl19}
& \alpha < 0, \, \,  \, \beta < 0, \, \, \, &{\rm Case~I, ~BB ~vector ~solitons,}   
\\
\label{sl20}
& \alpha > 0, \, \, \, \beta < 0, \, \, \, &{\rm Case~II, ~BD ~vector ~solitons,}   
\\
\label{sl21}
& \alpha > 0,\, \, \, \beta > 0, \, \, \, &{\rm Case~III, ~DB ~vector ~solitons,}   
\\
\label{sl22}
& \alpha < 0, \, \, \,  \beta > 0, \, \, \, &{\rm Case~IV, ~DD ~vector ~solitons \, .}   
\end{eqnarray}

\begin{figure}[htp]
    \centering
    \includegraphics[width=16cm]{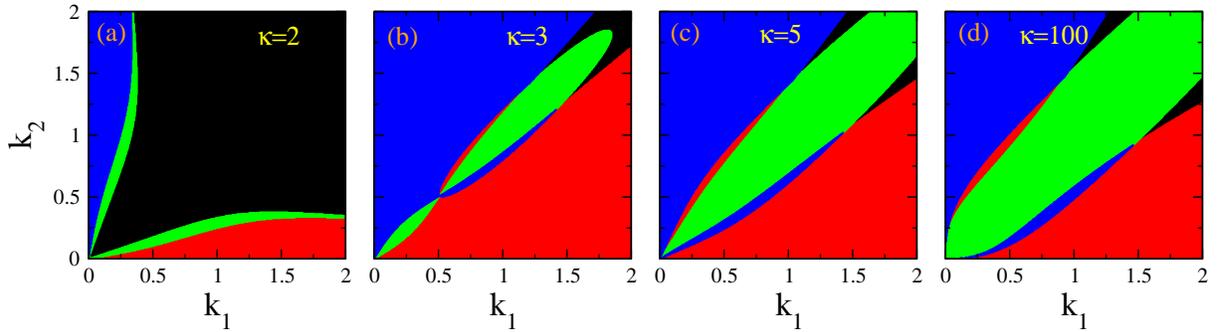}
    \caption{
    Existence areas for bright-bright (BB, black color), bright-dark (BD, red 
    color), dark-bright (DB, blue color), and dark-dark (DD, green color), on 
    the $k_1 - k_2$ parameter plane, for $A_1 =0.1$ and (a) $\kappa=2$; 
    (b) $\kappa=3$; (c) $\kappa=5$; (d) $\kappa=100$. 
    }
    \label{fig3}
\end{figure}
\begin{figure}[htp]
    \centering
    \includegraphics[width=16cm]{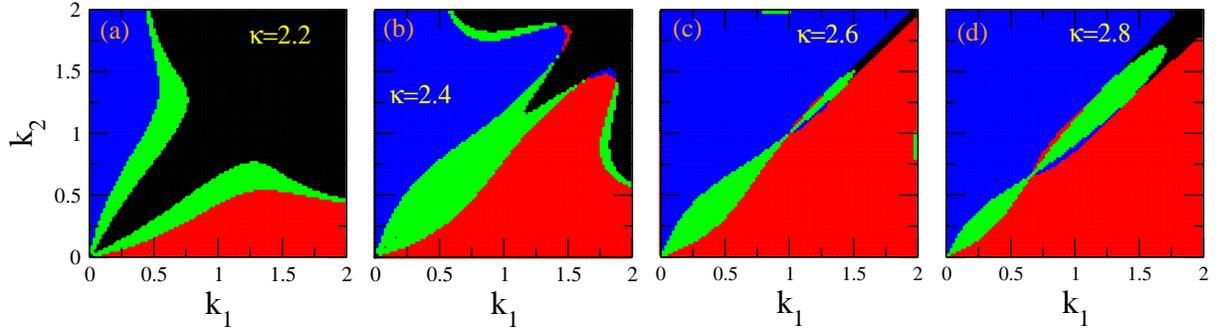}
    \caption{
    The same as Fig. \ref{fig3} for values of $\kappa$ between $2$ and $3$.
    Existence areas for bright-bright (BB, black color), bright-dark (BD, red 
    color), dark-bright (DB, blue color), and dark-dark (DD, green color), on 
    the $k_1 - k_2$ space, for $A_1 =0.1$, and (a) $\kappa=2.2$; (b) $\kappa=2.4$; 
    (c) $\kappa=2.6$; (d) $\kappa=2.8$. 
    }
    \label{fig4}
\end{figure}

In Fig. \ref{fig3}, we have identified on the $k_1 - k_2$ plane those areas in
which each type of vector soliton may exist, for several values of the spectral 
index $\kappa$ spanning a very wide range from $2$ to $100$ (for the latter, the 
kappa distribution has practically converged to a Maxwell-Boltzmann one). In
Fig. \ref{fig3}(a), the black color that corresponds to BB vector solitons is 
dominant, and occupies a large part of the plane except that for low $k_1$ and 
$k_2$ where BD (red color) and DB (blue color) vector solitons may exist. Note
that between the areas of existence of BB and DB, as well as between BB and BD,
DD vector solitons exist (green color) in two separate narrow areas. 
Interestingly, the latter are grown around the path where $Q_{12}$ and $Q_{21}$
are zero on the $k_1 - k_2$ plane. For $\kappa =3$ or greater, different 
patterns of existence of vector solitons appear (Fig. \ref{fig3}(b)), which 
typically consist of two large areas of existence of BD and DB vector solitons 
(red and blue color, respectively) while in between of these two there is a green 
area in which DD vector solitons exist. A small black area in which BB solitons 
may exist also appears at the upper right corner of the plane. This pattern 
appears for all larger values of $\kappa$, with only slight quantitative 
differences observed in Figs. \ref{fig3}(c)-\ref{fig3}(d)). Obviously, the 
strongest variation of the existence patterns of vector solitons on the 
$k_1 - k_2$ plane occurs between $\kappa =2$ and $\kappa =3$. In order to 
analyze this pattern variability, we present in Fig. \ref{fig4} a series of 
patterns for four (4) values of $\kappa$ between $2$ and $3$. Here we see how 
the dominant (black) area of BB vector solitons gradually shrinks against both 
the existence areas for BD and DB vector solitons (red and blue area, 
respectively).
\begin{figure}[htp]
    \centering
    \includegraphics[width=14cm]{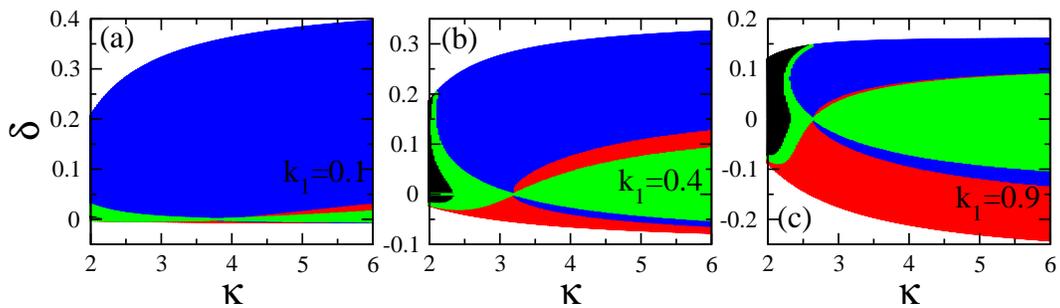}
    \caption{
   Existence areas for bright-bright (BB, black color), bright-dark (BD, red 
    color), dark-bright (DB, blue color), and dark-dark (DD, green color), as a 
    function of $\kappa$ and $\delta$ for $A_1 =0.1$ and (a) $k_1 =0.1$; 
    (b) $k_1 =0.4$; (c) $k_1 =0.9$. The values of $\delta$ are obtained by 
    varying the wavenumber $k_2$ from $0$ to $2$ and the corresponding fixed $k_1$. 
     }
    \label{fig44}
\end{figure}

It may be interesting to display the existence area for various types of vector solitons 
on a plane in which one of the coordinates is the ``walk-off'' parameter $\delta$.
In Fig. \ref{fig44}, these existence areas are shown on the $\kappa - \delta$
plane, for specific indicative values of the carrier wavenumbers $k_1$ and $k_2$. 
The color code for vector solitons is the same as that used in Figs. 
\ref{fig3} and \ref{fig4}. The walk-off parameter is calculated by setting $k_1$
to a fixed value, and then varying $k_2$ from $0$ to $2$. In Fig. \ref{fig44}(a),
dark-bright (DB, blue color) vector solitons are dominant in the whole $\kappa$ 
interval. Small areas in which dark-dark (DD, green color) and bright-dark (BD,
red color) however exist are visible for small positive values of $\delta$.
Much richer patterns appear in Figs. \ref{fig44}(b) and (c), for larger wavenumber 
$k_1$, in which all four
types of vector solitons may exist in substantial parts of the $\kappa - \delta$
plane. In both figures, bright-bright (BB, black color) vector solitons exist 
in small areas of the plane at low $\kappa$ ($\kappa < 2.5$).

\section{Illustrative Examples of Vector Solitons and their Parameters}

\begin{figure}[htp]
    \centering
    \includegraphics[width=12cm]{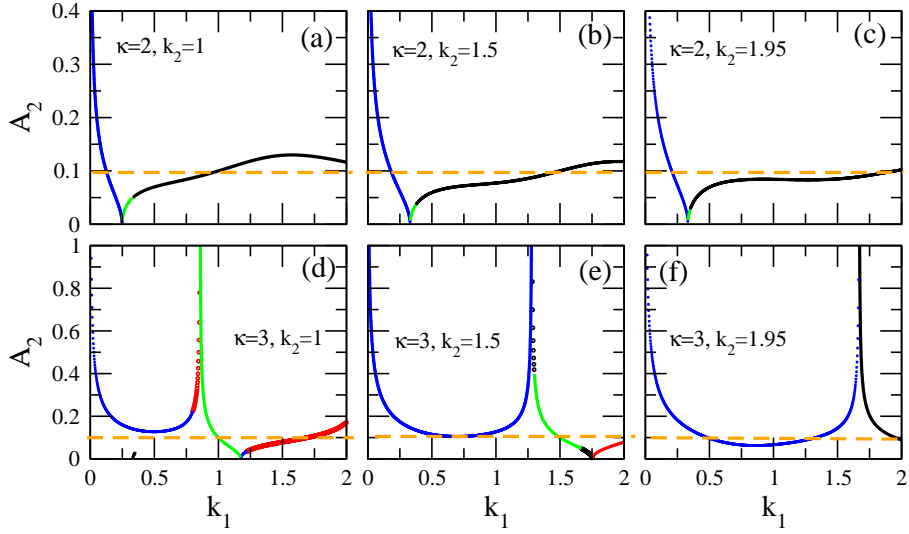}
    \caption{
    The amplitudes of the vector solitons components $A_1$ and $A_2$ as a function 
    of the wavenumber $k_1$, calculated from the first of Eqs. (\ref{sl17}). 
    The fixed amplitude $A_1 =0.1$ is indicated by the horizontal orange dashed line. 
    (a) $\kappa =2$, $k_2 =1$;    (b) $\kappa =2$, $k_2 =1.5$; 
    (c) $\kappa =2$, $k_2 =1.95$; (d) $\kappa =3$, $k_2 =1$;
    (e) $\kappa =3$, $k_2 =1.5$;  (f) $\kappa =3$, $k_2 =1.95$. 
    Note that the same color code as in Figs. \ref{fig3} and \ref{fig4} has 
    been adopted in all curves shown, i.e. black/red/blue/green color represents 
    values prescribing BB/BD/DB/DD vector solitons, respectively.
    }
    \label{fig5}
\end{figure}

It is tempting to investigate how the vector soliton characteristics vary when 
one of the wavenumbers of the carrier waves. e.g., $k_1$ varies (while the rest 
of the parameters as well as the amplitude $A_1$ remain fixed). 
In Fig. \ref{fig5}, the amplitude $A_2$ is plotted as a function of $k_1$ for 
two values of the $\kappa$, i.e., for $\kappa =2$ (upper panels) and $\kappa =3$ 
(lower panels), and for three different values of the wavenumber $k_2$ 
($k_2 =1$, $1.5$, $1.95$). For the upper panels, the three values of $k_2$ 
correspond to three ``cuts'' (sections) at $k_2 =1$, $1.5$, $1.95$ of the 
$k_1 - k_2$ plane in Fig. \ref{fig4}. As obvious in this figure, for all three 
values, DB type vector solitons occur for low $k_1$, while upon increasing $k_1$ 
they successively turn to DD and then BB soliton pairs.

Fig. \ref{fig5}(a)-(c) depicts the variation of the magnitude of (the amplitude) 
$A_2$ as well as its behavior when boundaries of areas of existence of different 
types of vector solitons are crossed. The three plots exhibit similar behavior: 
the amplitude $A_2$ takes very high values for low $k_1$ (DB vector solitons) which 
gradually decreases with higher $k_1$ and almost vanishes at the boundary between 
DB (blue) and DD (green) vector solitons (existence regions). 
As soon as this boundary is crossed, the amplitude $A_2$ increases again with 
increasing $k_1$, it reaches a maximum slightly above $A_1 =0.1$ and then decreases 
again slightly. Moreover, the second boundary crossing, i.e. between DD and BB 
vector solitons, is a smooth and continuous process as evident from the continuity 
of the green into the black part(s) of the plotted curve. interestingly, for 
$k_1 > 0.5$, the two amplitudes $A_1$ and $A_2$ are of the same order of magnitude.

The corresponding curves in the lower panels, obtained for $\kappa =3$, also 
exhibit a vanishing amplitude at the boundary between the areas of existence of 
DD to BD vector solitons. However, a different type of behavior is now witnessed, 
as the amplitude $A_2$ diverges at the boundary separating BD- from DD-type 
vector soliton regions (areas of existence): see Fig. \ref{fig5}(d).  
In the former case, the variation of the amplitude $A_2$ through the corresponding 
boundary is smooth, while in the latter it is not. In the same figure, one  also 
witnesses a smooth variation of $A_2$ through the boundary between DB (blue) and 
BD (red) vector solitons, as well as a divergence of the amplitude when crossing 
the boundary between the areas of existence of BD (red) and DD (green) vector 
solitons. Similar remarks can be made for Figs. \ref{fig5}(e) and (f). 
In the latter figure, only one boundary crossing is visible for the $k_1$ 
interval shown.

\begin{figure}[htp]
    \centering
    \includegraphics[width=12cm]{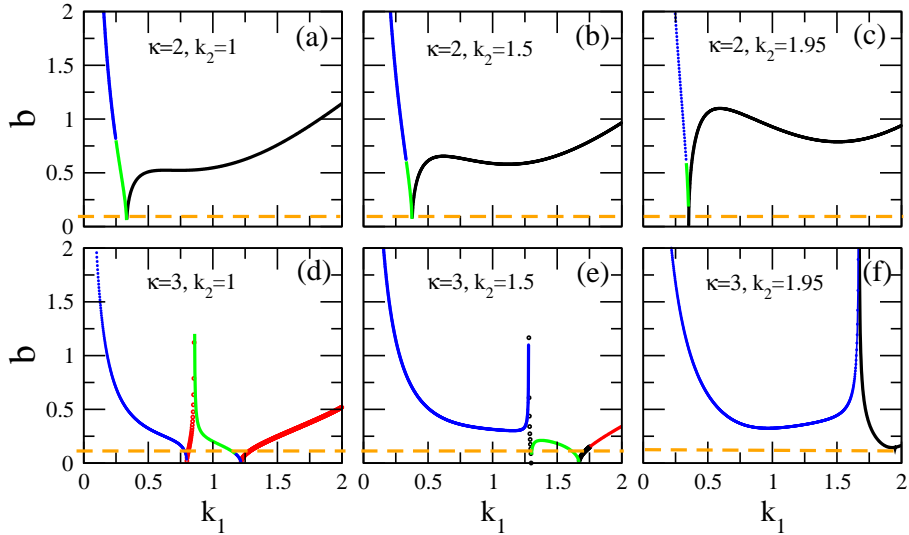}
    \caption{
    The width $b$ of the vector soliton components as a function of the 
    wavenumber $k_1$, calculated from the second of Eqs. (\ref{sl17}). 
    The fixed amplitude $A_1 =0.1$ is indicated by the horizontal orange dashed line. 
    (a) $\kappa =2$, $k_2 =1$;    (b) $\kappa =2$, $k_2 =1.5$; 
    (c) $\kappa =2$, $k_2 =1.95$; (d) $\kappa =3$, $k_2 =1$;
    (e) $\kappa =3$, $k_2 =1.5$;  (f) $\kappa =3$, $k_2 =1.95$. 
    Note that the same color code as in Figs. \ref{fig3} and \ref{fig4} has 
    been adopted in all curves shown, i.e. black/red/blue/green color represents  
    values prescribing BB/BD/DB/DD vector solitons, respectively.
    }
    \label{fig6}
\end{figure}

The (common) width $b$ of the vector soliton components associated to the 
amplitudes $A_2$ (Fig. \ref{fig5}) are presented in Fig. \ref{fig6}. The width 
$b$ in all the sub-panels in Fig. \ref{fig6} takes very high values at low $k_1$, 
up to $k_1 \simeq 0.1$, suggesting a very extended (spatially) -- i.e. little 
localized -- solution, but with high amplitude $A_2$. (Recall that $A_1$ is fixed 
in this figure.)

The behavior of $b$ with $k_1$ increasing beyond $0.1$ is then diversified for 
the two values of the spectral index shown: $\kappa =2$ (upper panels) and 
$\kappa =3$ (lower panels). Consider the former case first, i.e., that with 
$\kappa =2$. For $k_1$ increasing above $0.1$, the width $b$ decreases almost 
linearly but abruptly, passing smoothly through a boundary from DB (blue) to DD 
(green) vector soliton existence areas. At approximately $k_1 \simeq 0.3$, the 
width $b$ reaches a very low value of the order of $0.05$, while another boundary 
crossing between DD (green) to BB (black) vector solitons takes place at that 
point. For further increasing $k_1$, the width $b$ increases again, its exact 
behavior however this time depends on the selected value of the second wavenumber 
$k_2$. In Fig. \ref{fig6}(a), for $k_2 =1$, the width $b$ after a linear increase 
it saturates to a value that is approximately constant around $b \simeq 0.5$. 
Then, for $k_1$ increasing above $~0.8$, the width $b$ increases parabolically. 
For the two other values of $k_2$, i.e., for $k_2 =1.5$ and $k_2 =1.95$ shown 
in Figs. \ref{fig6}(b) and (c), respectively, the width $b$ as a function of 
$k_1$ does not form a plateau but instead it reaches a maximum, then a shallow 
minimum, and finally it start increasing parabolically upon further increasing 
$k_1$. Both the maximum and minimum values of $b$ are higher for $k_2 =1.95$, 
as compared to the case with $k_2 =1.5$.

\begin{figure}[htp]
    \centering
    \includegraphics[width=14cm]{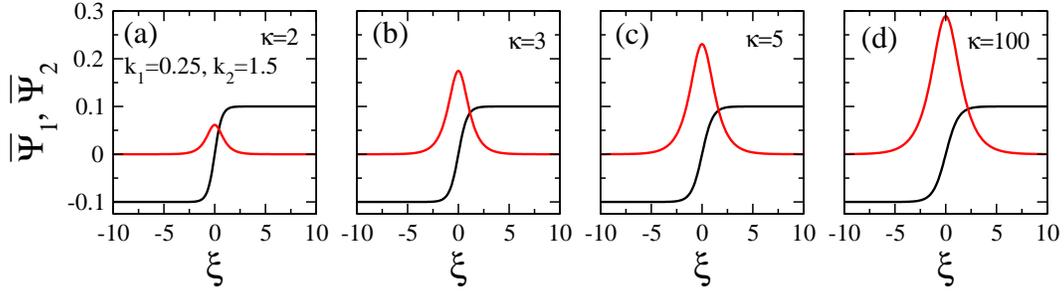}
    \caption{
    Variation of the vector soliton components as the spectral index $\kappa$
    takes the values (a) $\kappa =2$; (b) $\kappa =3$; (c) $\kappa =5$; 
    (d) $\kappa =100$, for a fixed point on the $k_1 - k_2$ plane, i.e., for 
    $k_1 =0.25$ and $k_2 =1.5$. 
    The envelops $\bar{\Psi}_1$ and $\bar{\Psi}_2$ are plotted as a function of $\xi$. 
    }
    \label{fig7}
\end{figure}
For $\kappa =3$, as shown in Figs. \ref{fig6}(d)-(f), the width $b$ of the 
vector solitons also decreases when $k_1$ increases further above $0.1$ but not
as abruptly as in the corresponding case with $\kappa =2$. In the $k_1$ interval 
shown in Figs. \ref{fig6}(d), (e), and (f), we observe $3$, $3$, and $1$ boundary 
crossings, respectively (for $k_2 =1$, $1.5$, and $1.95$). Specifically, for 
$k_2 =1$ (Fig. \ref{fig6}(d)), the first boundary crossing appears at about 
$k_1 \simeq 0.8$ where from DB (blue) vector solitons one transits into 
BD (red) ones smoothly (i.e. the curve representing the width $b$ is continuous) 
at very low values. The second crossing, from BD (red) to DD (green) type vector 
solitons occurs very close to the first one and is not smooth in this case, in 
the sense that $b$ diverges at the boundary. A third crossing occurs at 
$k_1 \simeq 1.25$ where again the transition of $b$ is smooth with very low 
values of $b$ at the crossing point. The latter crossing is from DD (green) to 
BD (red) vector solitons.  
For $k_2 =1.5$ (Fig. \ref{fig6}(e)), a crossing between existence areas from DB
(blue) to DD (green) vector solitons occurs at approximatelly $k_1 \simeq 1.25$
which is not smooth in $b$, since it diverges there. A second, smooth crossing 
occurs at around $k_1 \simeq 1.7$ in which $b$ has low values and separates 
existence areas of DD (green) and BB (black) vector solitons. For that value of 
$k_2$, BB (black) vector solitons exist only for a very short interval of $k_1$, 
and the third crossing in this figure occurs at approximately $k_1 \simeq 1.75$ 
between existence areas of BB (black) to BD (red) vector solitons. This crossing 
is smooth and continuous in the width $b$. For $k_2 =1.95$ (Fig. \ref{fig6}(f)), 
a single crossing occurs between existence areas of DB (blue) and BB (black) 
vector solitons. The behavior of $b$ as a function of $k_1$ has the same features 
as that of the amplitude $A_2$ shown in Fig. \ref{fig5}.

The divergence of the amplitude $A_2$ and the width $b$ for low $k_1$ in Figs. 
\ref{fig5} and \ref{fig6} is due to the corresponding divergence of the 
coefficient $Q_{11}$, which appears in the numerator of both $\alpha$ and $\beta$ 
in Eq. (\ref{sl05}). However, there is also a divergence at relatively large $k_1$  
for $\kappa =3$, signaling the transition from one type of vector soliton to 
another. Specifically, in Figs. \ref{fig5}(d) and \ref{fig6}(d) a divergence in 
$A_2$ and $b$ sets the boundary between a bright-dark (BD) and a dark-dark (DD) 
vector soliton at $k_1 =0.86$. That divergence occurs because both $\alpha$ and 
$\beta$ change sign through their denominator crossing the zero line. When this 
happens, a transition of a BD vector soliton (which exists for $\alpha >0$ and 
$\beta < 0$) to a DD vector soliton (which exists for $\alpha < 0$ and $\beta > 0$) 
takes place. (Note that the denominators of $\alpha$ and $\beta$ differ by an 
overall sign only.) Very close to the transition point, where the amplitude of 
the second component may be much larger than unity, the CNLS equations (\ref{eq22}) 
and (\ref{eq23}) are not expected to provide a valid description of vector solitons 
in our plasma fluid model.

\subsection{Extremely asymmetric waves emerging as vector soliton components}

Note that, with reference to Figs. \ref{fig5} and \ref{fig6}, there may exist
boundaries between existence areas of different type of vector solitons in which
the amplitude $\Psi_2$ of the envelope soliton component acquires large values, 
while its width $b$ acquires very low values simultaneously. Such a case can be
observed in Fig. \ref{fig6}(e), for a value of $k_1$ slightly above $1.25$ 
(green curves in Figs. \ref{fig5}(e) and \ref{fig6}(e)), for which DD vector 
solitons exist. For that $k_1 \simeq 1.3$, the width $b$ acquires very low 
values while at the same time the amplitude $A_2$ of the second vector soliton 
component acquires values around $A_2 \simeq 0.4$, which are four (4) times 
larger that the corresponding amplitude $A_1 =0.1$ (fixed) of the first vector 
soliton component. Such a highly localized and high-amplitude envelope soliton 
can be characterized as ``extremely asymmetric wave'', i.e. a large amplitude 
breather-like structure which co-exists (co-propagates) with an ordinary sister 
envelope soliton (pulse). Of course, the specific choice of values makes the 
components of the (DD in the described case) vector soliton highly a-symmetric. 
This possibility, as well as the stability and geometry of extremely asymmetric
vector soliton (pairs) will be analyzed in detail in a future a work.

\begin{figure}[htp]
    \centering
    \includegraphics[width=14cm]{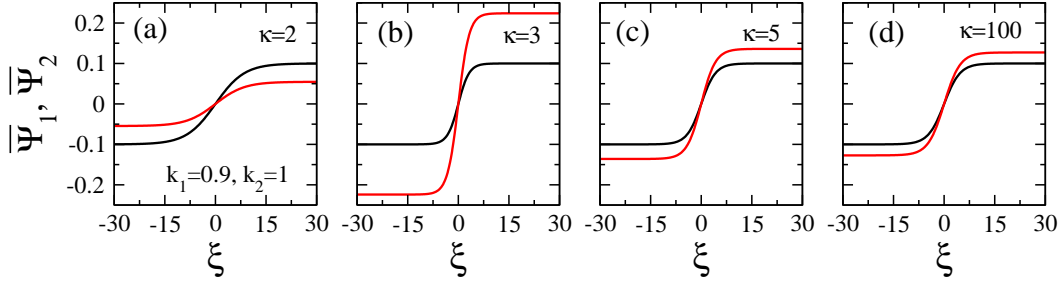}
    \caption{
    Same as Fig. \ref{fig7} but for $k_1 =0.9$ and $k_2 =1$. 
    }
    \label{fig8}
\end{figure}

\begin{figure}[htp]
    \centering
    \includegraphics[width=14cm]{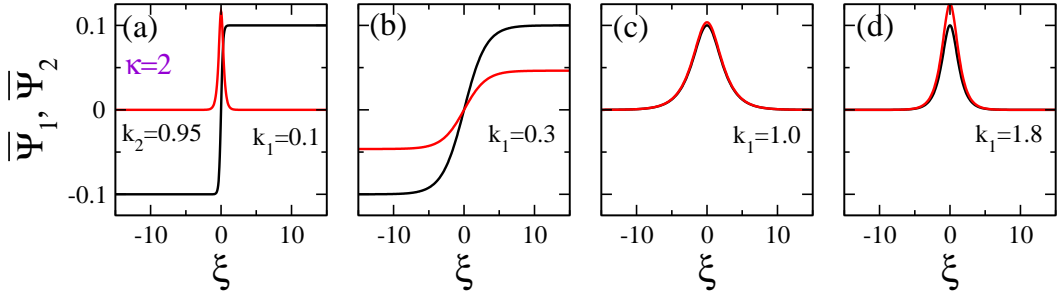}
    \caption{
    Vector solitons along a ``cut'' of the $k_1 - k_2$ plane in Fig. \ref{fig3}(a) 
    at $k_2 =0.95$ for $\kappa =2$ and (a) $k_1 =0.1$; (b) $k_1 =0.3$;
    (c) $k_1 =1$; (d) $k_1 =1.8$. We observe a DB vector soliton in (a), a DD 
    vector soliton in (b), and a BB vector solitons in (c)-(d).
    }
    \label{fig9}
\end{figure}

\begin{figure}[htp]
    \centering
    \includegraphics[width=14cm]{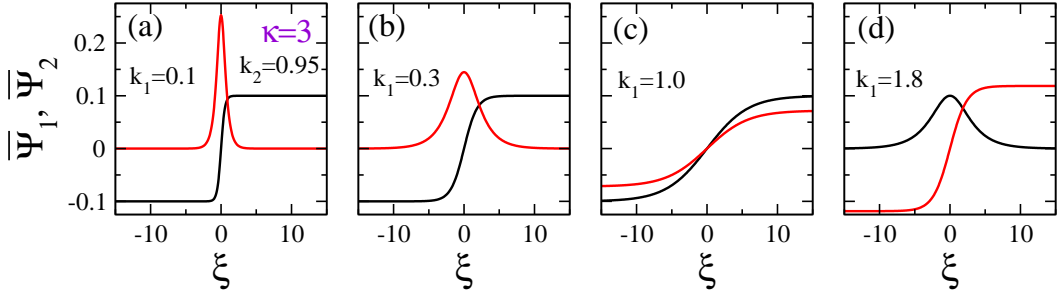}
    \caption{
    Vector solitons along a ``cut'' of the $k_1 - k_2$ plane in Fig. \ref{fig3}(b) 
    at $k_2 =0.95$ for $\kappa =3$ and (a) $k_1 =0.1$; (b) $k_1 =0.3$;
    (c) $k_1 =1$; (d) $k_1 =1.8$. 
    We observe a DB vector soliton in (a)-(b), a DD vector soliton in (c), and 
    a BD vector soliton in (d).
    }
    \label{fig10}
\end{figure}

In Figs. \ref{fig7}-\ref{fig10}, several illustrative examples of all the four
types of vector solitons which may exist in certain areas on the $k_1 - k_2$ 
plane, are provided for several values of the spectral index $\kappa$ and the 
wavenumbers $k_1$ and $k_2$. The first two of these figures, in particular, i.e., 
Fig. \ref{fig7} and Fig. \ref{fig8}, provide illustrative examples of DB and DD 
vector solitons, respectively, for $\kappa =2$, $3$, $5$, and $100$.
E.g., in Fig. \ref{fig7}, the values for the two wavenumbers were chosen to be 
$k_1 =0.25$ and $k_2 =1.5$ (for all subfigures), while $\kappa$ varies as shown 
on the figure. In this particular case, the values of $k_1$ and $k_2$ favor the 
existence of DB vector solitons for any $\kappa >2$. 
Note that the amplitude of the envelop soliton component (black curve), $\Psi_1$ , 
$A_1$, is fixed to $0.1$, and thus it is the same in Figs. \ref{fig7}(a) through 
(d), even though its width $b$ does change slightly its value. The second 
vector soliton component (red curve), $\Psi_2$, has a varying amplitude which 
increases considerably with increasing $\kappa$, as can be observed in 
Fig. \ref{fig7} (e.g., from $A_2 \simeq 0.06$ in Fig. \ref{fig7}(a) to 
$A_2 \simeq 0.29$ in Fig. \ref{fig7}(d).

Similarly, in Fig. \ref{fig8}, the same plots of vector solitons as those 
presented before in Fig. \ref{fig7} are shown but for a different $k_1$ and 
$k_2$ pair of values on the $k_1 - k_2$ plane, i.e., $k_1 = 0.9$ and $k_2 =1$.
for this choice of the $k_1$ and $k_2$ pair, we get DD vector solitons for all 
the values of the spectral index $\kappa$ considered. Of course the width of 
the envelopes $b$ and the amplitude of the vector soliton component $\Psi_2$,
$A_2$, change considerably with $\kappa$. Interestingly, the width $b$ of both
soliton components is rather large for an extreme (strongly non-Mawxellian) 
value $\kappa =2$ (Fig. \ref{fig8}(a)). Also, the amplitude $A_2$ is very low in 
this case, i.e., $A_2 =0.05$. In Fig. \ref{fig8}(b), however, for $\kappa =3$, 
the width $b$ decreases considerably, so that the soliton components become 
narrower and thus highly localized, while at the same time the amplitude $A_2$ 
increases to more than $0.22$. With further increasing $\kappa$, both $b$ and 
$A_2$ gradually and slowly decrease, while they tend to become equal to each 
other for very large values of $\kappa$, e.g., for $\kappa =100$ in Fig. \ref{fig8}(d).

The vector solitons shown in Figs. \ref{fig9} and \ref{fig10} are obtained for
spectral index $\kappa =2$ and $3$, respectively, and different pairs of 
wavenumbers $k_1$ and $k_2$, along $k_2 =0.95$ on the $k_1 - k_2$ plane. In 
Fig. \ref{fig9}(a), a DB vector soliton is shown for $k_1 =0.1$ and $k_2 =0.95$, 
which are both very narrow (small width $b$) and have roughly the same amplitude, 
i.e., $A_1 \simeq A_2 \simeq 0.1$. A DD vector soliton is shown in Fig. 
\ref{fig9}(b) for $k_1 =0.3$ and $k_2 =0.95$, where the amplitude $A_1$ is twice 
the amplitude $A_2$. In Figs. \ref{fig9}(c) and (d), two BB vector solitons are 
shown for $k_1 =1.0, k_2 =0.95$, and $k_1 =1.8, k_2 =0.95$, respectively. 
The two components of these vector solitons do not differ very much and, in one 
case (Fig. \ref{fig9}(d)), the two amplitudes $A_1$ and $A_2$ as well as their 
width $b$ are practically the same. This seems reasonable since the values of 
$k_1$ and $k_2$ in this case are very close together and thus close to the curve 
$k_1 =k_2$ on the $k_1 - k_2$ plane, where the CNLS system becomes nearly symmetric.

In Fig. \ref{fig10}, several vector solitons are shown for $\kappa =3$ and the 
same pairs of values of the wavenumbers $k_1$ and $k_2$ as those used in 
Fig. \ref{fig9}. Here, a narrow (small $b$) highly localized DB vector soliton 
is shown in Fig. \ref{fig10}(a), whose $\Psi_2$ component has a rather high 
amplitude $A_2 \simeq 0.3$, about three times the amplitude $A_1$.
In Fig. \ref{fig10}(b) we have another DB vector solitons whose components are 
however much wider than the previous one (large $b$) and $A_2$ is roughly 
$1.5 A_1$. In Fig. \ref{fig10}(c), we have a DD vector soliton whose components 
have similar amplitudes but they are rather wide (large $b$), Next, in
Fig. \ref{fig10}(d), we have two BD vector solitons, whose components exhibit 
comparable amplitude.

\begin{figure}[htp]
    \centering
    \includegraphics[width=12cm]{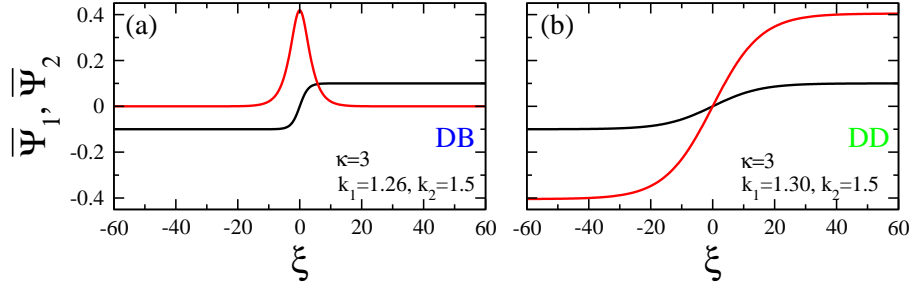}
    \caption{
    Extreme amplitude (asymmetric) vector solitons obtained around the 
    boundary of the area of existence of dark-bright (DB, blue color) and 
    dark-dark (DD, green color) in Fig. \ref{fig3}(b) for spectral index 
    $\kappa =3$, $k_2 =1.5$, and (a) $k_1 =1.26$; (b) $k_1 =1.30$. 
    In both plots, the amplitude of the second component of the vector soliton 
    (bright and dark, respectively) is significantly larger than that of the 
    first one.
    }
    \label{fig11}
\end{figure}

An illustrative example of vector solitons with an extreme amplitude component
is shown in Fig. \ref{fig11}. In that figure, two such vector solitons obtained 
for the same spectral index $\kappa =3$ and wavenumber of the second carrier wave 
$k_2 =1.5$ are shown for slightly different values of $k_1$. These parameter 
values correspond to points very close to the boundary of the area of existence 
of dark-bright (DB, blue color) and dark-dark (DD, green color) of 
Fig. \ref{fig3}(b) for $\kappa =3$. At these points, the second components of the 
existing vector solitons are expected from Fig. \ref{fig5}(e) to have a high 
amplitude as compared with the first one. In Fig. \ref{fig11}(a), for $k_1 =1.26$, 
in which the vector soliton is of the DB type, the amplitude of the second (bright) 
component which is shown in red color is about four times larger than that of the 
first one. The same remark holds for the amplitude of the second (dark) component 
shown in red color in Fig. \ref{fig11}(b), for $k_1 =1.30$, is again about four 
times larger than the first one. Note the both in Figs. \ref{fig11}(a) and 
\ref{fig11}(b) the first component of the corresponding vector soliton is shown 
in black color and it is of the dark type. Thus, in this particular case one may 
switch from a certain type of vector soliton with a large component to another type 
(i.e., bright to dark, in this case) by a small shift in one of the wavenumbers, 
thus crossing a boundary between adjacent areas in the existence diagram.  
Note that the extreme amplitude component is the one that changes type in this 
case, i.e. from bright in Fig. \ref{fig11}(a) to dark in Fig. \ref{fig11}(b). 
(Recall that the amplitude of the first component in all vector solitons 
presented here is numerically fixed to $A_1 =0.1$, for comparison and reference.) 

\begin{figure}[htp]
    \centering
    \includegraphics[width=12cm]{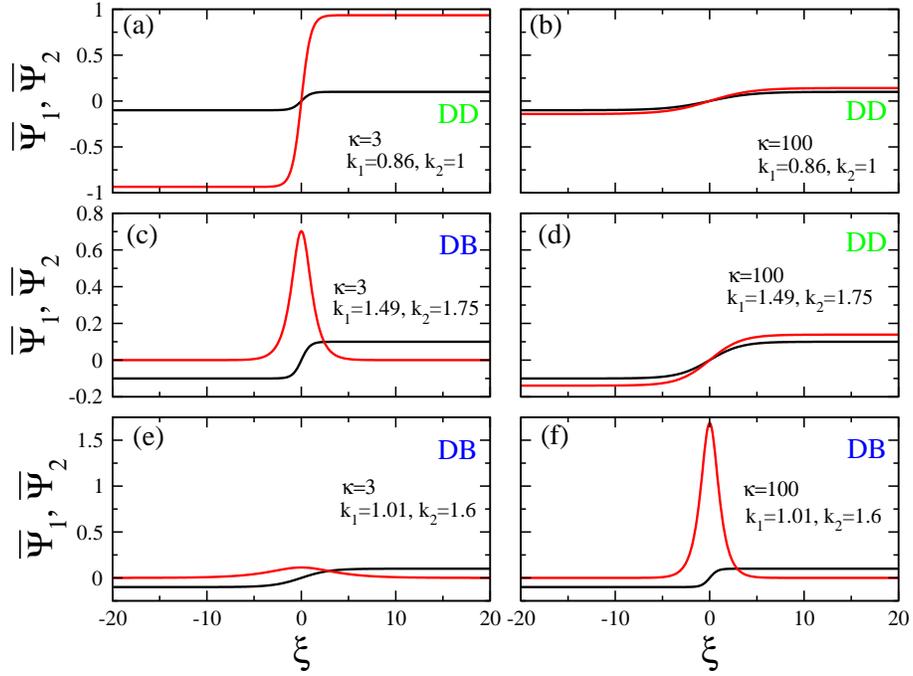}
    \caption{
    Comparison among representative vector soliton pairs obtained for small and 
    large values of the spectral index $\kappa$, i.e., for $\kappa =3$ (left panels; 
    strongly non-thermal distribution): 
    (a) $k_1 =0.86$, $k_2 =1$; 
    (c) $k_1 =1,49$, $k_2 =1.75$;
    (e) $k_1 =1.01$, $k_2 =1.6$,
    and $\kappa =100$ (right panels; quasi-Maxwellian): 
    (b) $k_1 =0.86$, $k_2 =$;
    (d) $k_1 =1.49$, $k_2 =1.75$;
    (f) $k_1 =1.01$, $k_2 =1.6$.
    The vertical scales are the same for all panels in each row, to facilitate 
    comparison. The horizontal scales are the same for all panels.
    }
    \label{fig12}
\end{figure}

Further examples of vector solitons with an extreme amplitude component are 
shown in Fig. \ref{fig12}, in which the effect of extreme variation of the 
spectral index $\kappa$ is illustrated. The two panels in each row are obtained 
for the same pairs of the wavenumbers $k_1$ and $k_2$ but for different $\kappa$. 
In all left panels, the value of spectral index $\kappa =3$ has been used, while 
in all right panels a large value of $\kappa$ is used ($\kappa =100$) for which 
the electron distribution practically coincides with a Maxwell-Boltzmann one. 
In Fig. \ref{fig12}(b), for $\kappa =100$, the selected values of $k_1$ and $k_2$ 
are such that the vector soliton is of the dark-dark (DD) type whose components 
have almost the same (relatively low) amplitude. However, as the spectral index 
is decreased to the relatively small value $\kappa =3$ in Fig. \ref{fig12}(a), 
the amplitude of the second component increases significantly with respect to 
that of the first one, although the the vector soliton type (DD) remains the same. 
Thus, by decreasing $\kappa$, the resulting increase in the suprathermal electron 
population provides the necessary energy for the emergence of a dark type extreme 
amplitude excitation shown in red color.

In the second row, i.e. in Figs. \ref{fig12}(c) and \ref{fig12}(d), we observe 
the emergence of an extreme amplitude \emph{bright} type component upon decreasing 
the value of $\kappa$, respectively. It appears that moving far from Maxwellian 
equilibrium results in energizing the electrons, that supply the energy required 
to excite a large amplitude ``breather'' type envelope structure seven (!) times 
higher than the sister component. A different trend is witnessed in 
Figs. \ref{fig12}(e) and \ref{fig12}(f). In this case, somehow counter-intuitively, 
a small-amplitude dark (wave 1)  / large amplitude bright (wave 2) vector soliton 
excited for $\kappa =100$ -- see \ref{fig12}(f) --  is in fact suppressed in 
amplitude, upon reducing the spectral index to $\kappa =3$ -- see \ref{fig12}(e) 
-- however without changing structural type (DB).
\begin{figure}[htp]
    \centering
    \includegraphics[width=7cm]{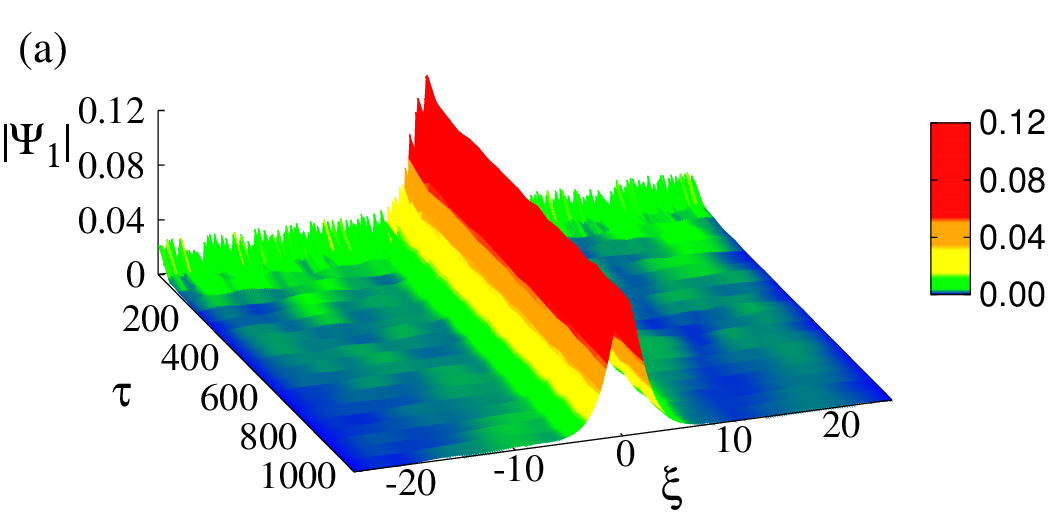} \hspace{3mm}
    \includegraphics[width=7cm]{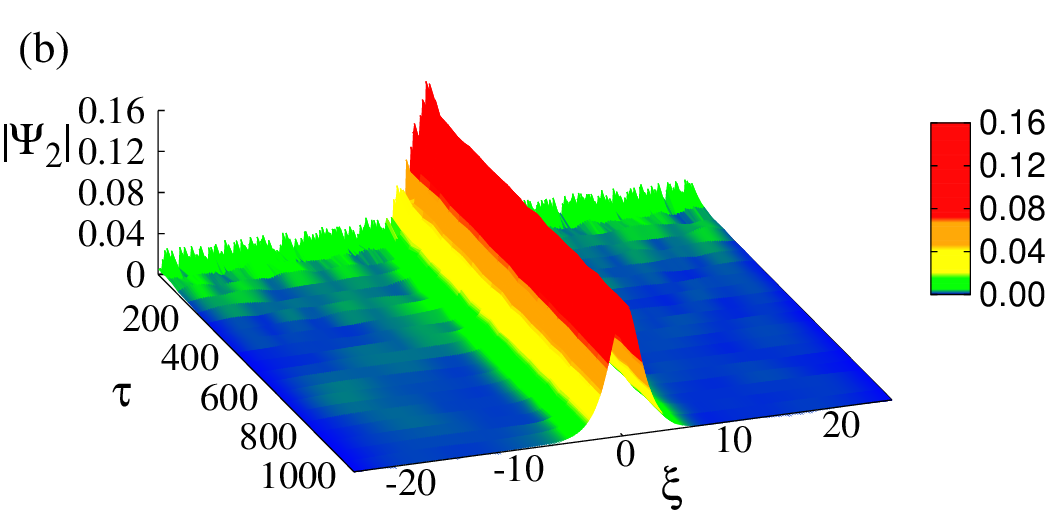} \\
    \includegraphics[width=7cm]{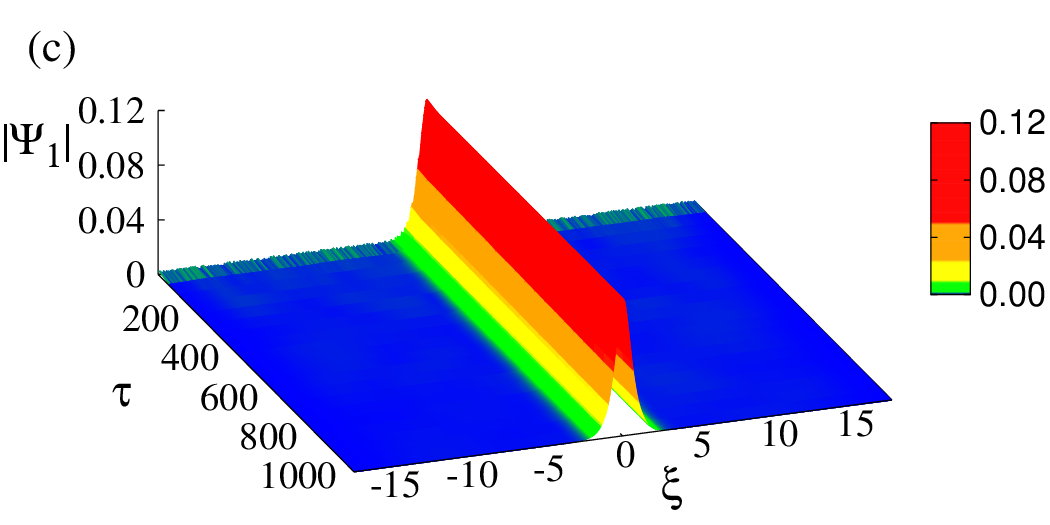} \hspace{3mm} 
    \includegraphics[width=7cm]{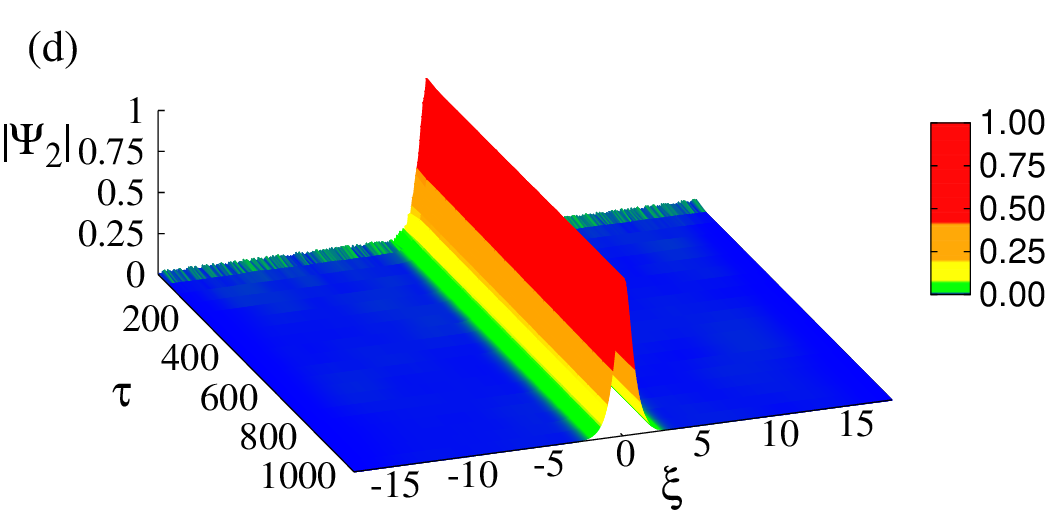}
    \caption{
    Temporal evolution of bright-bright vector solitons profiles obtained from 
    numerical simulations using Eqs. (\ref{eq22}) and (\ref{eq23}) with initial
    condition from Eq. (\ref{sl01}) to which small random noise is added.
    The profiles in (a) and (b) (resp. (c) and (d)) for $\bar{\Psi}_1$ and 
    $\bar{\Psi}_2$ are obtained for $k_1 =1.5$ and $k_2 =1$ (resp. $k_1 =1.673$ 
    and $k_2 =1.95$). The parameters of the vector soliton in (a) and (b) (resp.
    in (c) and (d)) are $A_1 =0.1$, $A_2 \simeq 0.129$, $b \simeq 0.759$ (resp.
    $A_1 =0.1$, $A_2 \simeq 0.964$, $b \simeq 2.13$).
    }
    \label{fig13}
\end{figure}

In order to address the stability of the obtained vector solitons, linear 
stability analysis corroborated by direct numerical simulations and/or (numerical)
spectral stability analysis should be used to obtain the parameter regimes 
which can support stable vector solitons. Further analytical approaches may 
involve methods applied, e.g., to birefringent optical fibers \cite{Kivshar1990}
or the application of the Vakitov-Kolokolov criterion \cite{Ostrovskaya1999}.
The stability analysis of the various types vector solitons in plasmas is 
certainly important and worth studying on its own right and it is a matter of 
future work. Here, stability is demonstrated using direct numerical simulations 
for two bright-bright (BB) vector solitons; the parameters are chosen so that the 
components of one of them have similar amplitude while the components of the 
other are highly asymmetric (e.g. the amplitude of the second component is much 
higher than that of the first).

These two cases are illustrated in Fig. \ref{fig13} in which (a) and (b) show 
the temporal evolution of the $|\bar{\Psi}_1|$ and $|\bar{\Psi}_2|$ profiles for 
the first set of parameters, while (c) and (d) those for the second set of 
parameters (see caption in Fig. \ref{fig13}). These simulations are initialized 
with the corresponding analytical solutions given in Eqs. (\ref{sl01}), to which 
small random noise was incorporated. In Figs. \ref{fig13}(a) and (b) that random 
noise is relatively large so that it is clearly visible for small $\tau$. 
At both ends of of the $\xi$ interval, dissipation has been added by hand to 
remove the excess energy introduced by the noise. In this way, the excess energy 
leaves the system and at large $\tau$ the profiles practically coincide with the 
analytical ones. We have also checked that at large $\tau$ the norms of the two 
components  $|\bar{\Psi}_1|^2$ and $|\bar{\Psi}_2|^2$ saturate to a constant 
value.

\section{Discussion and Conclusions}

We have considered the simultaneous propagation of a pair of (nonlinearly interacting) electrostatic 
wavepackets in a collisionless unmagnetized 
electron-ion plasma, from first principles. The wavepackets are not identical, 
in the sense that both their amplitudes and (carrier) wavenumbers $k_1$ and $k_2$ 
are allowed to differ. Adopting a Newell type multiple (time and space) scales 
technique, a pair of CNLS equations was derived. A standard non-magnetized plasma 
fluid model was adopted, for simplicity, comprising cold inertial ions evolving 
against an inertialess electron background. The electron population was assumed 
to obey a kappa-type distribution, which is characterized by the spectral index 
parameter $\kappa$, a situation often occurring in space plasmas (with typical 
values of $\kappa$ usually ranging from $2$ to $6$). As the kappa distribution 
diverges significantly from a Maxwell-Boltzmann distribution, the highly energetic 
(suprathermal) electron component results in significant modification of the 
modulated wavepackets characteristics and interactions thereof. 
We have investigated the modulational (in)stability profile of the coupled 
wavepacket pair, focusing on its dependence on the electron spectral index 
($\kappa$). We have shown that various types of vector solitons may exist in 
different areas on the $k_1 - k_2$ plane, while their shape depends on 
(the value of) $\kappa$. 
The strongest variation is observed in the interval of $\kappa$ from $2$ to $3$.

The six coefficients of the CNLS equations, i.e., the dispersion coefficients 
$P_j$, the nonlinearity coefficients $Q_{jj}$, and the nonlinear coupling
coefficients $Q_{ij}$ (with $i\neq j$), are given by complicated algebraic 
expressions (see Supplementary Information) as functions of the wavenumbers 
$k_1$ and $k_2$ and the spectral index $\kappa$.
For arbitrary values of $k_1$ and $k_2$ ($\ne k_1$), these coefficients do not 
possess any particular symmetry, hence the generalized system of CNLS equations 
thus obtained is most likely non-integrable, in the general case. 
(Obviously, the integrable Manakov case is recovered if $k_1 = k_2$.) 
The asymmetry of the CNLS equations does not prevent one from obtaining vector 
soliton solutions, as combinations of bright (B) and dark (D) envelopes 
structures, i.e. of BB, BD, DB or DD type, each of which will occur in 
particular areas on the $k_1 - k_2$ parameter plane.

The ``area of existence'' of these vector solitons exhibits strong variation 
with respect to $\kappa$, in particular for values of $\kappa$ between $2$ and $3$. For
$\kappa$ close to $2$, BB vector solitons exist in the largest part of the 
plane, while BD and DB vector solitons exist only for low $k_1$ and $k_2$,
respectively. Also, DD vector solitons exist in narrow areas between BB-BD and 
BB-DB vector solitons. However, for increasing $\kappa$, the area of existence 
of BD and DB vector solitons increases at the expense of the area of BB vector
solitons. At the same time, the area of existence of DD vector solitons also 
increases. For $\kappa$ greater than $3$, the pattern of the areas of existence
of the four vector solitons change only slightly with $\kappa$.

The vector soliton parameters, i.e., their amplitude and width, both of which 
have been calculated analytically, also vary significantly with varying $k_1$, 
$k_2$ and $\kappa$. It is interesting to see how the transition between different 
types of vector solitons occurs, upon varying one of these parameters, keeping 
the remaining two parameters fixed. As illustrated in the figures above, this 
transition between different types of vector solitons can be either smooth, 
or associated with a divergence of, say, the amplitude of one wave at the 
transition point, i.e. at the boundary separating areas of existence of different 
types of vector solitons. 

Of particular interest is the situation in which, close to a transition point 
where a soliton parameter diverges, the amplitude of one of the components may 
acquire extreme values, i.e. far exceeding its sister wave's amplitude, thus 
forming what could be characterized as an extreme amplitude wave (component) 
pair. These highly asymmetric vector solitons are a peculiarity which is attributed 
to the general asymmetry of the CNLS equations. The investigation of the stability
of these vector solitons using semi-analytic and numerical methods will be a subject
of future work.

Focusing on the role of suprathermal electrons, our investigation has shown that 
the spectral index $\kappa$ affects significantly the modulational (in)stability 
profile of the CNLS system as well as the characteristic of vector soliton types 
that may be sustained in the plasma. For smaller $\kappa$, i.e., as the electron 
distribution deviates from the Maxwellian one, the areas on the parameter planes 
in which modulational instability appears become larger, while at the same time 
the growth rate becomes higher in those areas (i.e., enhancing the  instability). 
Concerning vector solitons, a variation in the value of $\kappa$ modifies the 
existence diagram in parameter space, in which different types of vector solitons 
may occur. The most prominent variation occurs below $\kappa =3$, where DB, BD, 
and DD vector solitons exist in substantial areas of the $k_1 - k_2$ plane, down 
to $\kappa =2$, where BB vector solitons become dominant.

The existence of all four types of vector solitons on a parameter plane involving
the spectral index $\kappa$ and the walk-off parameter $\delta$ was also illustrated 
and discussed. Notably, for the particular plasma model considered here, the 
parameter $\delta$ takes low values (in fact, the lower  $\kappa$, the smaller the 
range of $\delta$), which are thus not expected to affect the formation or the stability 
of vector solitons.

The formation of solitary waves/solitons is a phenomenon that commonly occurs 
in  space plasmas, e.g., in the solar wind and in planetary magnetospheres; 
cf. observations of electrostatic solitary waves by the Cluster satellites 
\cite{Graham2015,Lakhina2021} (and references therein). As one example, 
electrostatic solitary waves have been observed in the Earth’s magnetopause 
\cite{Trines2007,Stasiewicz2004}, and their theoretical interpretation  
requires resorting to multicomponent plasma fluid models \cite{Ji2014}. 
Remarkably, envelope structures (breathers, rogue waves) modeled by the NLS 
equation have been realized in laboratory plasmas, little more than a decade 
ago \cite{Bailung2011}.

Based on earlier considerations, where modulational instability has been proposed 
as an intermediate stage between amplitude modulation of a Stokes wave and 
higher-order effects leading to rogue wave formation \cite{neo1}, we anticipate 
that the creation of extreme amplitude soliton-pair structures predicted by our 
model may provide an effective framework as a precursor towards freak wave 
occurrence in relation with electtostatic plasma modes.

Our work aims at providing a platform for modeling solitons/solitary waves in 
space plasmas, where modulated envelope pairs may emerge from two or more 
interacting nonlinear waves. In a wider context, our results will be valuable 
in other disciplines where wavepackets may propagate in nonlinear dispersive 
media, including -- but not being limited to -- hydrodynamics, nonlinear (fiber) 
optics and telecommunications (signal transmission via optical pulses), 
to mention a few. 



\section*{Acknowledgements}
Authors IK and NL gratefully acknowledge financial support from Khalifa 
University (United Arab Emirates) via the 
project CIRA-2021-064 (8474000412).
IK gratefully acknowledges financial support from KU via the project 
FSU-2021-012 (8474000352) as well as from KU Space and Planetary Science Center,   
via grant No. KU-SPSC-8474000336. 

This work was completed during a two-semester long research visit by author IK 
to the Physics Department, National and Kapodistrian University of Athens, Greece. 
During the same period, author IK also held an Adjunct Researcher status at the 
Hellenic Space Center, Greece. The hospitality of both hosts is warmly acknowledged.

\medskip

\section*{Author contributions statement}
N.L. carried out the algebraic work, contributed to the methodology, 
software development, and numerical analysis. 
G.P.V. contributed to the methodology and reviewed the manuscript.
D.J.F. contributed to the concept and design, and reviewed the manuscript. 
I.K. contributed to the problem conception, project design, launch and  methodology, 
and reviewed the manuscript.

\section*{Competing Interests}
The authors declare no competing interests.

\section*{Data Availability}
The datasets generated during and/or analysed during the current study are 
available from the corresponding author upon reasonable request.

\end{document}